\documentclass{article}
\usepackage{geometry}
\usepackage{graphicx}
\begin{document}

\title{Epidemic Threshold for the SIRS Model on the Networks} 

\author{M. Ali Saif\\
Department of Physics\\
Faculty of Education
University of Amran
Amran,Yemen\\
masali73@gmail.com}
\maketitle
\begin{abstract}
We study the phase transition from the persistence phase to the extinction phase for the SIRS (susceptible/ infected/ refractory/ susceptible) model of diseases spreading on the networks. We derive an analytical expression of the probability for the descendants nodes to re-infect their ancestors nodes. We find that, in the case of the recovery time $\tau_R$ is larger than the infection time $\tau_I$, the infection will flow directionally from the ancestors to the descendants however, the descendants will not able to reinfect their ancestors during their infection time. This behavior leads us to deduce that, for this case and when the infection rate $\lambda$ is high enough in such that, any infected node on the network infects all of its neighbors during its infection time, SIRS model on the network evolves to extinction state, where all the nodes on the network become susceptible. Moreover, we assert that, in order to the infection occurs repeatedly inside the network, the loops on the network are necessary, which means the clustering coefficient will play an important role for this model. Hence, unlike the other models such as SIS model and SIR model, SIRS model has a two critical threshold  which separate the persistence phase from the extinction phase when $\tau_I<\tau_R$. That means, for fixed values of $\tau_I$ and $\tau_R$ there are a two critical points for infection rate $\lambda_1$ and  $\lambda_2$, where epidemic persists in between of those two points. We confirm those results numerically by the simulation of regular one dimensional SIRS system.

\end{abstract}


\section{Introduction}
The dynamics of infectious diseases propagation within a mean filed framework or networking frame has been attracted considerable attention recently \cite{andr,kermack,anderson,landa,kuperman,pastor,gade,roy,yan,rozh}. Mean field approach for this problem assumes that, populations are not spatially distributed so that they are fully mixed by means an infective individuals are equally likely to spread the disease to any other member of the population or subpopulation to which they belong \cite{andr}. With these zero dimensional models it has been possible to establish many important epidemiological results, including the existence of a population threshold values for the spread of disease \cite{anderson} and the vaccination levels required for eradication \cite{andr,kermack}, and the effect of stochastic fluctuations on the modulation of an epidemic situation \cite{landa}.
A second approach assumes spatially distribution of population in a lattice such that the variables at each node represent the state of an individual \cite{kuperman,pastor,gade,roy,yan,rozh}. An epidemiological model has been investigated on many kinds of networks such as regular networks \cite{kuperman,roy}, small world networks \cite{kuperman,gade,gam}, scale-free networks \cite{sat,yan}, random networks \cite{liu} and directed networks \cite{cong}. 

In this work, we aim to find the epidemic threshold which separates the persistence phase, where the infection can spread throughout a population, from the extinction phase, where the infection dies out for SIRS model on networks. Epidemic critical threshold has been derived for models such as, SIR model on random networks \cite{pastor,new}, SIS model on random networks \cite{pars}, SIS model on correlated complex networks \cite{bog} and the epidemic model described by a birth-death process \cite{dyk}. For the SIRS model on the regular 2D network, it was observed that, there are two critical thresholds separate the persistence phase from the extinction phase, i. e. for a fixed value of infection time $\tau_I$ the epidemic will be persistent when infection rate $\lambda$ is between a minimal critical infection rate $\lambda_1$ and a maximal critical infection rate $\lambda_2$ \cite{quan}. In this work, we support those results analytically and numerically for SIRS model on a regular of 1D network.
 
\section{Model Description}
SIRS epidemic model on the networks is defined as follows \cite{kuperman,gade,yan}:  
if we consider a one dimensional lattice of $N$ nodes. Each node connected to its $k$ neighbors. The nodes can exist in one stage of three state, susceptible $(S)$, infected $(I)$ and refractory $(R)$. 
Susceptible node can pass to
the infected state through contagion by an infected one.
Infected node pass to the refractory state after an
infection time $\tau_I$. Refractory nodes return to the
susceptible state after a recovery time $\tau_R$. The contagion
is possible only during the $S$ phase, and only by a
$I$ node. During the $R$ phase, the nodes are immune
and do not infect. The system evolves with discrete time
steps. Each node in the network is characterized by a
time counter $\tau_i(t)= 0, 1, ..., \tau_I + \tau_R\equiv \tau_0$ , describing its
phase in the cycle of the disease. The epidemiological
state $\pi_i$ of the node $(S, I, or R)$ depends on the phase
in the following way:
\begin{eqnarray}
\pi_i(t)&=S & \mbox{if} \tau_i =   0,\nonumber\\
\pi_i(t)&=I & \mbox{if} \tau_i \in (1,\tau_I),\nonumber\\
\pi_i(t)&=R & \mbox{if} \tau_i \in (\tau_I +1,\tau_0)
\end{eqnarray}
The state of a node in the next step depends on its
current phase in the cycle. A susceptible node stays as such, at
$\tau= 0$, until it becomes infected. Once infected, it goes
(deterministically) over a cycle that lasts $\tau_0$ time steps.
During the first $\tau_I$ time steps, it is infected and can potentially
transmit the disease to a susceptible neighbor.
During the last $\tau_R$ time steps of the cycle, it remains in
state $R$, immune and not contagious. After the cycle is
complete, it returns to the susceptible state. This model is usually called SIRS model. 

\section{Model Analysis}
In the following, we consider a synchronous update of a system of $N$ nodes on a network, in which each node is connected to $k$ neighbors. Since the infection can spread
repeatedly within the nodes on the network for SIRS model, which means, the disease can transfer from the infected node (say $i$) to its neighbors $j_1,...,j_k$, and these neighbors also can reinfect that node $i$. Here, we are going to derive an expression for the probability of the descendants $j_1,...,j_k$ to reinfect their ancestor $i$ for SIRS model, this probability is different from the probability of ancestor to infect its descendants. Roni Parshani et. al. has been calculated that probability for SIS model on random networks \cite{cohen}. Here, on the same manner, we are going to calculate that probability for SIRS model. We assume that, at each time step an infected node infects each of its neighbors with probability $\lambda$. Infected nodes remain as such for $\tau_I$ time steps, after which they become immune for $\tau_R$ time steps. The probability $T$ for the infected node $i$ to infect one of its neighbors during its infection time is \cite{new,pars,cohen}:
\begin{eqnarray}
T=\left[1-(1-\lambda)^{\tau_I}\right]
\end{eqnarray}
When the node $i$ infects one of its neighbors (say $j_1$), the node $j_1$ can only reinfect $i$ after $i$ has became susceptible again. To calculate the probability $\pi$ of that descendant $j_1$ to reinfect its ancestor $i$, we assume the node $i$ has been infected for $t$ time steps before infecting $j_1$. Since the infection time is $\tau_I$ time steps and recovery time is $\tau_R$ time steps, $i$ needs for $(\tau_I+\tau_R -t)$ time steps after infecting $j_1$ to become susceptible again. Therefore, the total time in which $j_1$ can infect $i$ is $\tau_I-(\tau_I+\tau_R -t) = t-\tau_R$. Thus, it is clear that, if $t\leq\tau_R$, the node $i$ will reach $S$ state, whereas the node $j_1$ will have reached $R$ state. In this case, it is impossible for the descendant $j_1$ to reinfect its ancestor $i$, then when $t\leq\tau_R$. 
\begin{eqnarray}
\pi=0 
\end{eqnarray}

However when $t>\tau_R$, the probability $\pi$ is given by conditioning on $t$ \cite{cohen}:
\begin{eqnarray}
\pi=\sum_{t=\tau_{R}+1}^{\tau_I} \frac{(1-\lambda)^{t-\tau_{R}-1}  \lambda}{1-(1-\lambda)^{\tau_I}}\left[1-(1-\lambda)^{t-\tau_{R}} \right] \nonumber\\
  = \frac{\left[1-(1-\lambda)^{\tau_I-\tau_R}\right]\left[1-(1-\lambda)^{\tau_I-\tau_R+1}\right]}{(2-\lambda)\left[1-(1-\lambda)^{\tau_I}\right]}
\end{eqnarray}    

\subsection{Case of $\tau_R\geq\tau_I$}
Consider the situation when the node $i$ infects anyone of its neighbors (say $j_1$) during the time which it is  $t\leq \tau_R$, according to the Eq. 3, the node $j_1$ will not be able to reinfect the node $i$ during its infection time, where $\tau_R\geq\tau_I$. For such case, the maximum value of $t$ is $t=\tau_I$, which is less than $\tau_R$, that means the infected nodes can not reinfect their ancestors during their infection time. Thus, for SIRS model and, when $\tau_R\geq\tau_I$, the infection will flow only in one direction, from ancestors to descendant. Moreover, the neighbors of any node who were infected on the time at which that node was infected, will not be able to reinfect that node. This result leads us to infer the following two points in which, SIRS model on the networks evolves to absorbing state under the condition of  $\tau_R\geq\tau_I$. 

The first point concerns to the effects of the loops inside the network. To clarify this point, we consider a network which is without loops, e. g. treelike network. For simplicity, we also assume the infection starts at a single node $i$ on that network. The node $i$ can infect anyone of its neighbors $j_1,j_2,...j_k$ with probability $\lambda$ at each time step, also those infected neighbors to the node $i$ can infect some of their neighbors, however they can not reinfect the node $i$ where $\tau_R\geq\tau_I$. It is clear that for such case, each link connects two nodes on the network will transmit the infection only in one direction, which means, the infection will
spread in a directional way away from the first infected node toward the network boundaries. Thus, SIRS model on such kind of networks will evolve to absorbing state where, all nodes become susceptible, that will happen for any values of infection time and infection rate. We conclude that, {\it the existing of the loops on the networks is necessary to spread the infection frequently within the nodes on the networks.}

The second point relates to the effects of the values of the parameters $\tau_I$ and $\lambda$. To show that effect, suppose the node $j_1$ and node $j_2$ are neighbors, in which the node $j_1$ has been infected for $t$ time steps before infecting $j_1$.
Then, the possibility to the node $j_2$ to infect the node $j_1$ at the time when the node $j_1$ becomes susceptible again, will be zero unless $t$ is greater than $\tau_R$. 
Now, if we consider the extremist case, in which each infected node on the network, infects all of its neighbors during its first infection period. If we also suppose the infection starts at a single node $i$ on the network, in which, that node infects all of its neighbors $j_1,...,j_k$ during its infection time. Where $\tau_R\geq\tau_I$, the nodes $j_1,...,j_k$ will not be able to reinfect the node $i$ during their infection time. Consecutively, the nodes $j_1,...,j_k$ can transmit the disease to their neighbors but these infected neighbors also can not to reinfect their ancestors nodes. Moreover, the maximum value of $t$ for any two connected nodes will not exceed the value of $\tau_I$, which means they also can not infect each other. Therefore, the infection in this case will propagate away from the center of infection (first infected node) toward the network boundaries, only in one direction. This means, each node on the network will be infected only a once. Therefore, such system will evolve to disease-free equilibrium in which all nodes become susceptible. Generally, {\it we can say that,  for SIRS model on the networks, and when $\tau_R\geq\tau_I $, if each infectious node on the network, infects all of its neighbors during its first stage of infection, SIRS model on that network evolves to an absorbing state where all the nodes on the network become susceptible}.We speculate that, such behavior will be correct for SIRS model on any kind of networks.

 Here arises a question, for this model on the networks,
 how the infection can spread repeatedly among the nodes on the networks? 
 The answer is that, firstly, the network should contain loops, which means clustering coefficient will play an important role in this model. 
 Secondly, there must be some neighbors on the network which they have been infected at different times, and have that difference in time greater than $\tau_R$. 
 For that to happen, it is necessary to have some infected nodes on the network so that these nodes  do not infect all their neighbors during the period of their illness. These uninfected nodes, they must be infected by another infected neighbor at a later time. In this case, we may find some neighbors in the network who have been infected by a time difference $t>\tau_R$. These neighbors with $t>\tau_R$, will induce the second period of infection. Here, it is clear that, the first reinfected node will be reinfected through a loop of nodes. Hence, the values of $\tau_I$ and $\lambda$ will play a crucial role for SIRS model on the networks. However, that does not mean the value of $\tau_R$ has no effect,
  it is manifest that, for a fixed value of $\tau_I$, the possibility of getting a pair of neighbors with $t>\tau_R$ increases as the value of $\Delta\tau=\tau_R-\tau_I$ decreases,  i. e. when the value of $\Delta\tau$ is minimum, there is a better chance to get some neighbors on the network with $t>\tau_R$.
  
It is clear from the Eq. (2) that, the probability for an infected node to infect its neighbors increases as the infection time or the infection rate increases. An infected node infects more numbers of its neighbors during its infection time as the value of $\lambda$ or $\tau_I$ becomes greater. Therefore, there must be a minimum value of $\tau_I$ as the value of $\lambda$  is kept fixed, at which each infected node can infect all of its neighbors during its first infection period. Similarly, if we fix $\tau_I$ and increase $\lambda$, there must be a minimum value of $\lambda$ at which each infected node can infect all of its neighbors during its first infection period. Thus, there is an upper critical threshold which separates the persistence phase from the extinction phase for SIRS model on networks.  
 From Eq. (2) we can find a rough relationship between the infection time and probability of infection at which the critical threshold exists. For small values of $\lambda$, we have $T\approx  \lambda\tau_I$ \cite{new}. Therefore, if $\lambda \tau_I= 1$ that means, each infected node will infect all of its neighbors during its first infection period. Such system eventually reaches an extinction state. Therefor, the critical threshold which separates the extinction phase from the persistence phase is
 \begin{eqnarray}
 \tau_I\equiv 1/\lambda 
 \end{eqnarray}
However, when $\tau_I< 1/\lambda$ there is a possibility for SIRS model to settles down to coexistent state with susceptible and infected nodes. 

The lower critical threshold, we can deduce it from the basic reproductive ratio $R_0$. Which defined as the average  number of secondary infections caused by a single infected node during its
infected time when it placed into a full
susceptible nodes. Hence, if $R_0 <1$, the introduced infected will recover without being able to replace themselves by new infections. However, for SIRS model on a network, the expected number of 
susceptible neighbors that a node has, when it just becomes infected is given by $\kappa-1$, where 	 
the total expected number of neighbors is $\kappa$ (where $\kappa\equiv \left\langle k^2\right\rangle/\left\langle k\right\rangle$ is the branching factor) \cite{pars}, and one of them must be excluded as the infected 	 
parent from which the current node descended. Then, the mean number of secondary infections 	 
per infected node is thus $(\kappa-1)T$ \cite{pars}. The infection will die out if each infected node does not spawn on average at least one replacement so, for a very large network, the critical threshold is given by the relation $(\kappa-1)T = 1$ \cite{pars}. For small values of $\lambda$ the epidemic threshold is:
\begin{eqnarray}
\tau_I\equiv 1/(\kappa-1)\lambda. 
\end{eqnarray}

This equation and Eq. (5) show that, the region on the phase space of the parameters ($\lambda$,$\tau_I$) at which the disease will persist, shrinks as the value of $\kappa-1$ decreases. It is clear that, Eq. (6) will collapse on Eq. (5) when $\kappa=2$. Hence, a persistence phase will not exist when $\kappa\leq 2$, which is the critical threshold for percolation \cite{pars,cohen}.

\begin{figure}
\includegraphics[width=70mm,height=60mm]{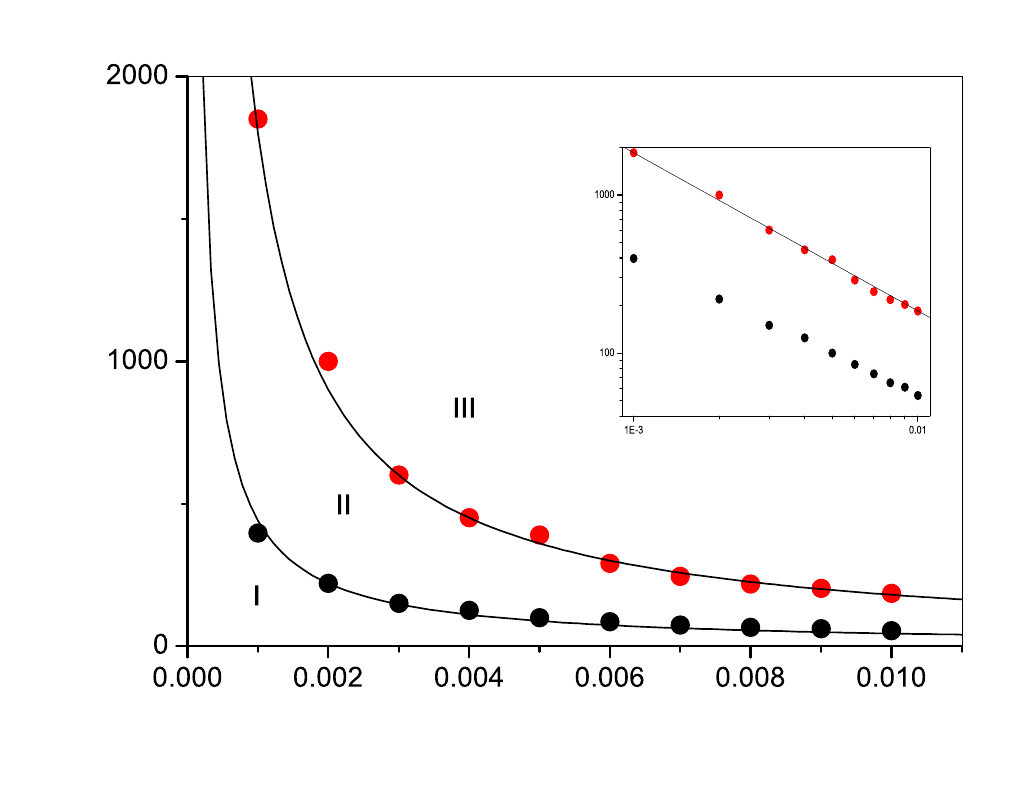}
\caption{(color on line)Phase diagram of infection time $\tau_I$ as function of infection probability $\lambda$ for regular 1D lattice with $k=3$ and total population $N=10000$. For each point, we set $\tau_R=\tau_I +1$ and, wait for $10000$ time steps. We averaged over $10$ independent runs. Inset shows double-logarithmic plot.}	
\end{figure}

Fig. 1, shows simulation phase diagrams (dotted curves) and analytical phase diagram (solid lines) for regular 1D system when $k=3$ and $N=10000$. Figure shows clearly the two critical thresholds. The region II correspond to persistence phase however, the regions I and III correspond to extinction phases. We fit upper solid curve with the relation 
$\tau_I= a \lambda^b$ where $a=1.85$ and $b=-1.0\pm 0.001$ for best fit. It is clear that, the simulation value of $b$ is in good agreement with the analytical expression Eq. (5).  Lower solid curve is fitted with the relation 
$\tau_I= a \lambda^b/(\kappa-1)$ where $\kappa=6$. For best fit, we find that $a=4.2$ and $b=-0.90\pm 0.01$. For this case, the simulation value of $b$ slightly diverts from the analytical value.

\subsection{Case of $\tau_R<\tau_I$}
In this case SIRS model will not differ much from SIS model, where we will restore SIS system when $\tau_{R}=0$ Eq.(4). For this case, lifetime of the disease is greater than the recovery time. Therefore, even if every infected node infects all their neighbors during its infection time, there is a possibility to get some nodes which has been infected with a time difference so that $t>\tau_R$. The probability for these nodes to reinfect their ancestors will be given by Eq. (4). Its clear that, this probability  increases as the value of $\Delta\tau$ increases. Hence, for this case, the upper critical threshold will not exist.	  
 
\section{Infection through the loop of nodes}
we expect that as SIRS model has two critical thresholds, then between of those critical thresholds there must be a value of $\lambda$ (for fixed value of $\tau_I$) at which this system will saturate to stable state with a highest infected nodes. To find an approximate relationship of that value, we estimate here the probability of infection through a loop of nodes. We guess that, when the probability of occurring the infection through a loop be maximum, SIRS model will saturate to a maximum infected nodes. Triangles are the smallest loops that connect three nodes. The probability of infection to occur through a triangle of nodes is:
\begin{eqnarray}
P=T^3(1-T) 
\end{eqnarray}

\begin{figure}
\includegraphics[width=80mm,height=40mm]{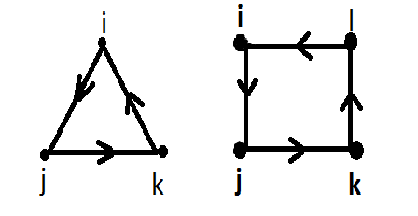}
\caption{Triangular loop and Square loop.}	
\end{figure}

 Fig.2 shows the triangular and square loop in which for triangular loop, the node $i$ infects its neighbor $j$ with probability T and, does not infect its neighbor $k$ with probability (1-T). The infected node $j$ now, infects its neighbor $k$ with probability T, consequently the node $k$ infects the node $i$ with probability T. Hence, probability for the node $i$ be infected through a triangular loop will be given by Eq. (7). This probability has two zero points at $T=0$ and $T=1$ , it has a peak at $T=\frac{3}{4}$ which is $\lambda=\frac{3}{(4\tau_I)}$  when $\lambda$ is small.
 
Probability for any node be infected through a square loop will be $T^4(1-T)$. For a loop of five nodes, that probability is $T^5(1-T)$, and so on for other loops. In Fig. 3, we show the probability of infection through different kinds of loops. Figure shows that, the probability of occurring the infection through a loop, it decreases as a loop be larger. 
\begin{figure}
\includegraphics[width=70mm,height=60mm]{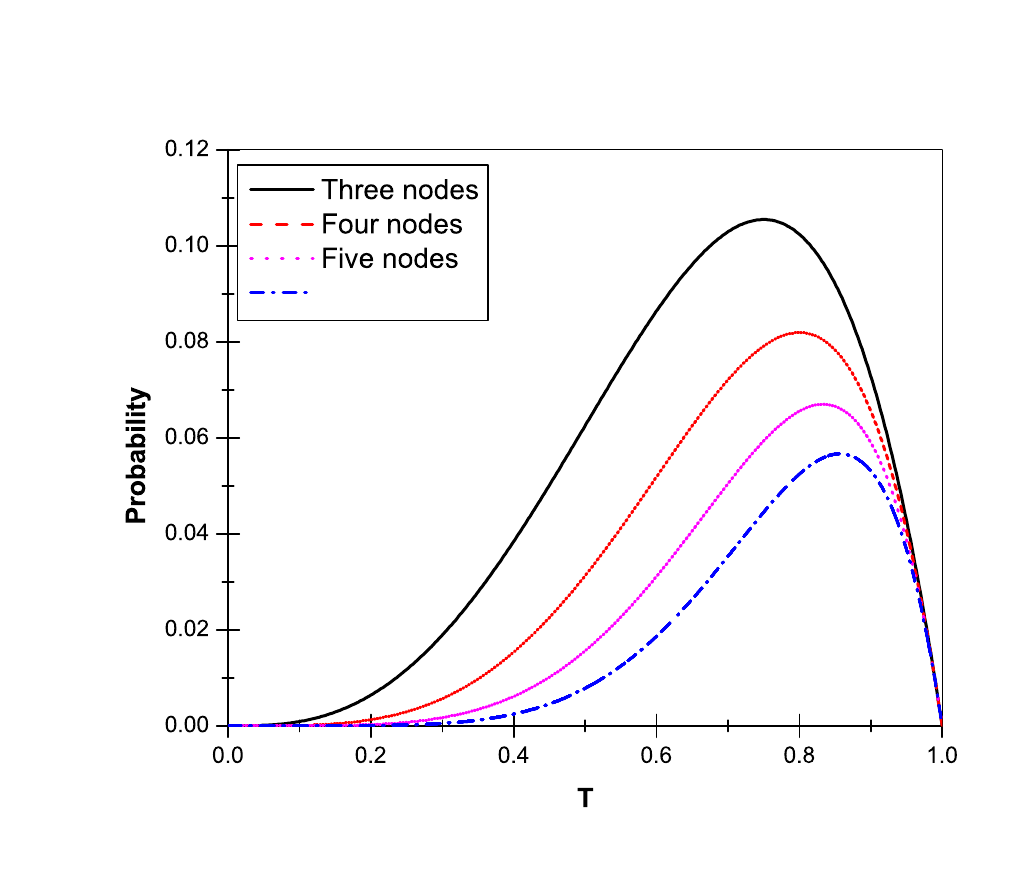}
\caption{Probability of occurring the infection through a loop of nodes as function of T, for different kinds of loops.}	
\end{figure}

Fig. 4 depicts the steady state of density of infection nodes as function of $\lambda$, for fixed values of $\tau_I=11$ and $\tau_R=12$. Simulations were performed 
in networks of size up to $N=10^5$, averaging over at least $100$ different realizations of the network. As we expected previously, the system shows a peak, at the value of $\lambda= 0.08\pm 0.01$ . This peak coincides approximately  with the peak at which the infection through the loops, is maximum. Which is $\lambda = 0.068$, for three nodes, $\lambda= 0.073$ for four nodes and $\lambda=0.076$ for five nodes. Figure shows also that, this system has two critical points, upper critical point at $\lambda=0.21\pm 0.002$, whereas lower critical point at $\lambda=0.052\pm 0.002$. 
\begin{figure}
  \includegraphics[width=70mm,height=60mm]{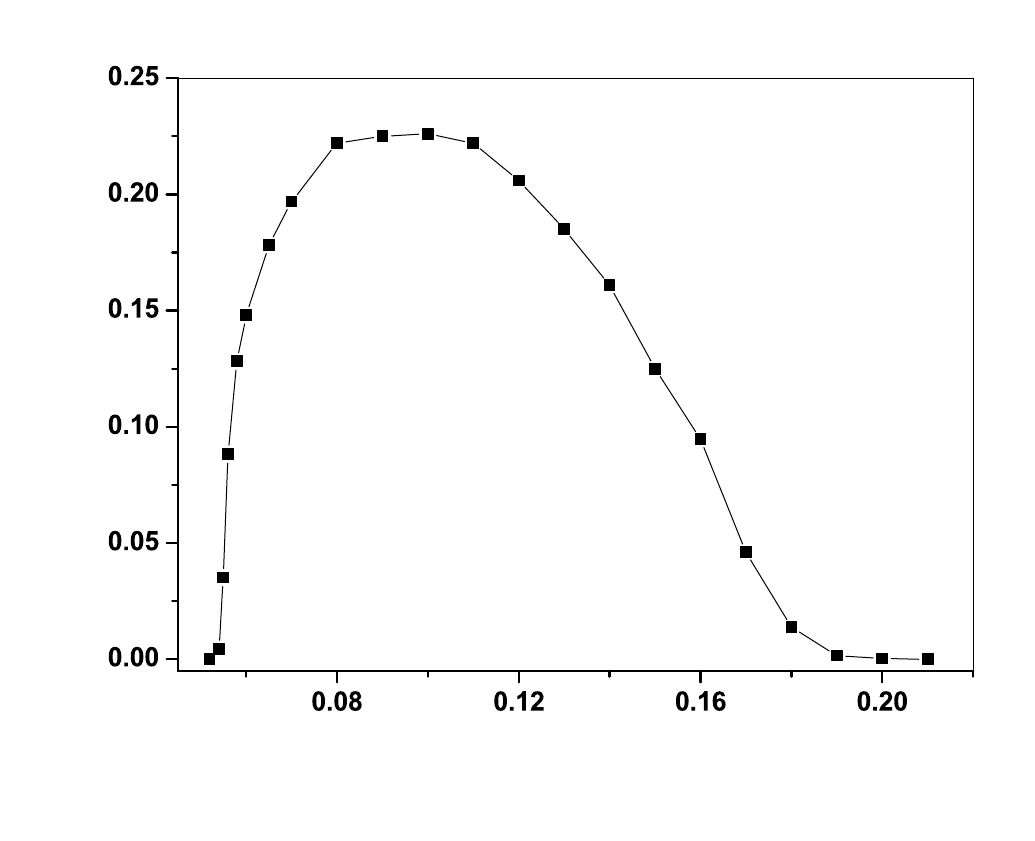}
	\caption{Steady state of density of infection nodes as function of infection probability $\lambda$ for regular 1D network with $k=3$, when $\tau_I=11$ and $\tau_R=12$.}	
\end{figure}

\section{Conclusion}
We have studied the spreading of  infectious diseases on the networks for SIRS model. We infer that, the probability that descendants will reinfect their ancestors will be zero, when $\tau_R\geq\tau_I$. Therefore, when the infection time or infection probability or both be high enough, such that, each infectious node on the network infects all of its neighbors during its first infection time, the nodes will pass only within one infection period. Such system evolves to an extinction state. On the other hand, if infection time or infection probability or both was low enough such that, each infectious node on the network can not infect on the average, one of its neighbors during its infection time, that system also evolves to an extinction state. Therefore, when $\tau_R\geq\tau_I$, SIRS on the networks has two critical thresholds separate the persistence phase from extinction phase. Epidemic only persists in between of these critical thresholds. However, when $\tau_R<\tau_I$, upper critical threshold will not exist. By numerical simulation of SIRS model on regular of 1D system, we created phase space and found epidemic thresholds.

\end{document}